\documentclass[prl,aps,showpacs,amsmath,amssymb,floatfix]{revtex4}

\usepackage[dvips]{graphicx} 
\usepackage{bm} 

\begin{document}

\title{Collective oscillations of a trapped Fermi gas  near a Feshbach
resonance}
\author{Sandro Stringari}
\affiliation{Dipartimento di Fisica, Universit\`a di Trento and BEC-INFM, I-38050 Povo, Italy}
  
\date{\today}

\begin{abstract} The frequencies of the collective oscillations of a
harmonically trapped Fermi gas  interacting with large  scattering lengths are
calculated at zero temperature using   hydrodynamic theory. Different regimes
are considered, including the molecular Bose-Einstein condensate   and the
unitarity limit for collisions. We show that the frequency of the radial
compressional mode in an elongated trap exhibits a pronounced non  monotonous
dependence on  the scattering length,  reflecting the role of the interactions
in the equation of state.
 
\end{abstract}


\maketitle

The recent realization \cite{D,G,M} of a Bose-Einstein condensate of pairs of Fermi atoms (molecular BEC) 
near a Fesbach resonance is opening new exciting perspectives in the investigation of highly correlated many-body systems including the investigation of the long sought BCS-BEC crossover \cite{randeria}. An important feature of these configurations is  the hydrodynamic nature 
of their macroscopic dynamics \cite{thomas} which, at zero temperature, is a general consequence of superfluidity.   Though the observation of the hydrodynamic behaviour cannot be used as a direct proof of
superfluidity  \cite{note}, its 
 careful investigation can provide unique information on the equation of state, thereby checking the predictions of many-body theories beyond the mean field picture. 

The purpose of this  letter is to show that the study of the collective oscillations in a Fermi gas close to a Feshbach resonance is well suited to this purpose, their frequencies exhibiting a non trivial dependence on the
scattering length in regimes now accessible experimentally \cite{grimm}.

Let us first consider the case of large and positive atomic scattering lengths close to a Feshbach resonance. This regime is associated with the occurrence of a weakly bound molecular state and consequently, for sufficently low temperatures and low densities,  one expects the formation of a Bose-Einstein condensate of molecules.   If the molecular gas parameter $n_Ma_M^3$ is sufficiently  small ($n_M$ and $a_M$ are, respectively, the molecular density and the molecule-molecule scattering length) the  molecular Bose-Einstein condensate is described by Gross-Pitaevskii theory.  In the presence of harmonic trapping one finds  that in the Thomas-Fermi limit  the frequencies of the collective oscillations do not depend on the interaction coupling constant, but only on the value of the axial and radial trapping frequencies
$\omega_z$ and $\omega_{\perp}$ \cite{sandro96}.  In the following we will discuss the case of elongated configurations ($\omega_z \ll \omega_{\perp}$) 
which are easily produced experimentally and where one can obtain simple analytic results at zero temperature. The coupled monopole-quadrupole  $m=0$ modes ($m$ is the third component of angular momentum) oscillate with frequencies \cite{sandro96} $\omega_{rad}=2\omega_{\perp}$ and $\omega_{ax}=\sqrt{5/2}\omega_{z}$  along
the radial and axial directions respectively.  These modes have been already observed in traditional Bose-Einstein condensates (see, for example, \cite{mit98,ens}). First measurements of $\omega_{ax}$ in a molecular condensate have been also reported \cite{G}. When the molecular gas parameter $n_Ma_M^3$ is no longer small the above results are modified by the presence of beyond mean field effects in the equation of state 
accounted for by the   expansion 
\begin{equation}
\mu_M = g_Mn_M\left(1+ {32\over 3\sqrt\pi}\sqrt{n_Ma^3_M}\right)
\label{mu}
\end{equation}
for   the molecular chemical potential. Here $g_M= 2\hbar^2 a_M/M$ is the molecule-molecule
interaction coupling constant and  $M$ is the atomic mass.  In terms of  the atomic chemical potential $\mu$ and density $n$ 
one has $\mu_M= 2\mu$ and  $n_M= n/2$. The relation between the molecule-molecule scattering length $a_M$ and
the atomic scattering length $a$ has been  recently  calculated  in \cite{petrov} and is given by $a_M=0.6 a$.  
 The collective frequencies predicted by the new equation of state   in the presence of harmonic trapping can be calculated using hydrodynamic 
 theory. The result for the $m=0$ radial frequency is \cite{PS}: 
\begin{equation}
\omega_{rad} = 2\omega_{\perp}\left[1+ {105\sqrt\pi \over 256 }\sqrt{n_M(0)a_M^3}\right] \; .
\label{omegarad}
\end{equation}
The relative correction to the mean field value $\sqrt{5/2}\omega_z$ of the $m=0$ axial frequency $\omega_{ax}$ turns out to be  a factor $6$ smaller, suggesting that the radial  mode is more suited to explore the interaction effects in the equation of state \cite{m2}. 
In the above equation $n_M(0)$ is the density of the molecular condensate evaluated in the center of the trap. 
It is useful to express the molecular gas parameter in terms of the relevant atomic  parameters. By using the Thomas-Fermi value for the central density
 one finds
\begin{equation}
n_M(0)a_M^3 = 0.06 \left(N^{1/6}{a\over a_{ho}}\right)^{12/5}
\label{na3}
\end{equation}
where $N$ is the total number of atoms   and $a_{ho}= \sqrt{\hbar/M\bar{\omega}}$ is the atomic oscillator length relative to the geometrical 
average $\bar{\omega}$ of the  trapping frequencies. 
When the scattering length $a$ is further increased and  becomes comparable or even larger 
than the average distance $d \sim n^{-1/3}$ between particles the molecular description is no longer applicable. Eventually,  when 
$a \gg d$ the system reaches the so called unitarity limit for collisions where the equation of state  does not depend any more on the actual value of the scattering length, nor on its sign. Provided the other relevant lengths associated with the two-body interaction, like the range of the potential, are smaller than $d$,  
the equation of state is expected to exhibit a universal behaviour \cite{heiselberg}.    Considerations of dimensionality suggest that at zero temperature  the chemical potential should have the same density dependence 
\begin{equation}
\mu = (1+\beta) {\hbar^2 \over 2m}(6 \pi^2)^{2/3}n^{2/3}
\label{beta}
\end{equation}
as in the ideal Fermi gas,
the parameter $\beta$ being a universal coefficient accounting for  quantum and dynamical correlations. At present the most reliable theoretical estimate of $\beta$  for a two component Fermi  system is  $\beta \sim -0.56$  obtained in \cite{pandha} through a  correlated basis function calculation. Actually the value of $\beta$ is not relevant for the determination of the collective frequency \cite{beta} which depends only on the power law coefficient characterizing  the density dependence of the chemical potential. If $\mu \propto n^{2/3}$  hydrodynamic theory predicts, for an elongated trap,  the values 
$\omega_{rad} = \sqrt{10/ 3}\omega_{\perp}$ and  $\omega_{ax} = \sqrt{12 / 5}\omega_{z}$ \cite{minguzzi}. The same results hold for  a classical gas in the hydrodynamic regime \cite{griffin} where, at constant entropy,  the chemical potential is also proportional to  $n^{2/3}$.  

It is worth noticing that the value predicted for  $\omega_{rad}$ is lower than the value $2\omega_{\perp}$ holding in the molecular BEC regime, revealing that  the frequency of the radial compression mode  first increases as a function of $a$, as predicted by (\ref{omegarad}),  and then decreases to reach the value $\sqrt{10/3}\omega_{\perp}$ in the unitarity limit (see  the Figure).  

Let us finally discuss the  region of negative scattering lengths, corresponding to the BCS regime of superfluidity.
When the gas parameter $n\mid a\mid^3$ is much smaller than unity  the chemical
potential still exhibits  a $2/3$ power law dependence on the density and
one  consequently expects to find again the result $\omega_{rad}= \sqrt{10/3}
\omega_\perp$ in the hydrodynamic regime. However, differently from the case
of positive scattering lengths,  in this limit  the critical temperature for
superfluidity becomes very small, and the superfluid phase is likely much more
difficult to reach.  Mean field corrections to the equation of state can be also easily calculated and yield
the expansion 
\begin{equation}
\mu = {\hbar^2 \over 2m}(6 \pi^2)^{2/3}n^{2/3} + {1\over 2}gn
\label{mug}
\end{equation}
for the chemical potential where $g=4\pi \hbar^2a/M$. By treating the interaction as a small perturbation  
 one finds the following result for  the 
frequency of the radial compression mode:
\begin{equation}
\omega_{rad} = \sqrt{10 \over 3} \omega_{\perp}\left(1 + {3\over 20}{E_{int}\over E_{ho}} \right) 
\label{a<0}
\end{equation}
where $E_{int} = (g/4)\int d{\bf r}n^2$ and $E_{ho}=\int d{\bf r}V_{ho}n$ are, respectively, the interaction and oscillator energies. In terms of the dimensionless combination $N^{1/6}a/a_{ho}$ one has $E_{int}/E_{ho}=0.5 N^{1/6}a/a_{ho}$ \cite{kf}. 
The mean field  correction in (\ref{a<0}) is a factor $4/5$ smaller than the analogous term calculated for the monopole mode in the case of isotropic trapping \cite{vichi}.
  The interaction correction in $\omega_{rad}$ is negative, reflecting the attractive nature of the force, so that  $\omega_{rad}$ first decreases  as a function of $\mid a\mid$ and then increases to reach the value  $\sqrt{10/3}\omega_{\perp}$
in the unitarity limit (see the Figure).   

The determination of $\omega_{rad}$ in the intermediate regimes  requires the full knowledge of the equation of state, through a consistent many-body calculation. In the Figure we have drawn a schematic interpolation (dot-dashed line) 
between the  asymptotic behaviours (\ref{omegarad}) and (\ref{a<0}) through the resonance. The curve  explicitly  points out the oscillatory behaviour of the frequency.

In conclusion we have shown that the frequency of the radial compression mode
in an ultracold trapped Fermi gas exhibits a non trivial dependence on the
scattering length near a Feshbach resonance, reflecting the occurrence of important interaction effects 
in the equation of state. These effects characterize  regimes of high physical interest, now available experimentally.  

Stimulating discussions with R. Grimm are acknowledged.

\begin{figure}
\includegraphics[width=8.5cm]{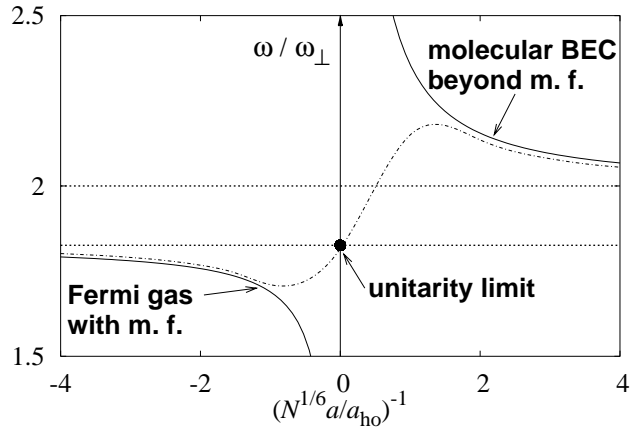}
\caption{\label{fig:?}Radial compressional frequency in an elongated trapped
Fermi gas close to a Feshbach resonance as a function of the dimensionless
parameter $(N^{1/6}a/a_{ho})^{-1}$. The mean field theory in the molecular BEC
regime predicts  $\omega/\omega_{\perp}=2$, while in the unitarity
limit one expects $\omega/\omega_{\perp}=\sqrt{10/3}\sim 1.83$. The full line for $a>0$ corresponds to the expansion (\ref{omegarad}) accounting for  beyond mean field (m.f.)
corrections in the molecular condensate with $a_M=0.6 a$, while the full line for $a<0$  corresponds to the expansion (\ref{a<0})
accounting for the mean field  correction in the BCS phase. The dot-dashed line is a schematic interpolation between the two asymptotic regimes.}
\end{figure}


\begin{thebibliography}{99}

  
\bibitem{D}
M. Greiner, C. A. Regal, and D. S. Jin, Nature 426, 537 (2003).

\bibitem{G}
S. Jochim, M. Bartenstein, A. Altmeyer, G. Hendl, S. Riedl, C. Chin, J. Hecker Denschlag and  R. Grimm, Science, {\bf 302}, 2103 (2003).

\bibitem{M}
M.W. Zwierlein, C.A. Stan, C.H. Schunck, S.M.F. Raupach, S. Gupta, Z. Hadzibabic, and W. Ketterle, Phys. Rev. Lett. {\bf 91}, 210402 (2003)
Phys. Rev. Lett. {\bf 91}, 210402 (2003).

\bibitem{randeria} See, for example, M. Randeria, in {\it Bose-Einstein Condensation}, edited by A. Griffin, D. Snoke and S. Stringari (Cambridge University Press,
Cambridge, England, 1995)

\bibitem{thomas} The first evidence of
hydrodynamic behaviour of a cold Fermi gas was provided by the observation of strong anisotropy effects in the expansion  after release of the trap,
K.M. O'Hara et al., Science {\bf 298}, 2179 (2002).

\bibitem{note} Also the normal phase of these strongly interacting systems can 
 exhibit  hydrodynamic  behaviour due to thermal  collisions.

\bibitem{grimm} Another important test of the equation of state is provided 
by the study of the radii of the trapped cloud, R. Grimm et al., to be published.

\bibitem{sandro96} S.~Stringari, Phys. Rev. Lett. \textbf{77}, 2360 (1996).

\bibitem{mit98} D.M. Stampern-Kurn, H.-J. Miesner, S. Inouye, M.R. Andrews, and W. Ketterle,  Phys. Rev. Lett. {\bf 81}, 500 (1998).

\bibitem{ens} F. Chevy, V. Bretin, P. Rosenbusch, K.W. Madison, and J. Dalibard,  Phys. Rev. Lett. {\bf 88} 250402 (2002).

\bibitem{PS} L. Pitaevskii and S. Stringari, Phys. Rev. Lett. {\bf 81}, 4541; E. Braaten and J. Pearson, Phys. Rev. Lett. {\bf 82}, 255 (1999).

\bibitem{m2} In the hydrodynamic regime the frequencies of the $m=\pm 2$  and $m=\pm 1$ quadrupole  modes are instead  independent of the equation of state and given by $\sqrt2 \omega_{\perp}$ and $\sqrt{\omega^2_z + \omega^2_{\perp}}$ respectively.
 
\bibitem{petrov} D.S. Petrov, C. Salomon and G. Shlyapnikov, cond-mat/0309010.

\bibitem{heiselberg} H. Heiselberg, Phys. Rev. A {\bf 63}, 043606 (2001).

\bibitem{pandha} J. Carlson, S.Y. Chang, V.R. Pandharipande and K. E. Schmidt, Phys. Rev. Lett. {\bf 91}, 050401 (2003).

\bibitem{beta} The value of $\beta$ is instead important to determine the size of the trapped cloud as well as its release energy.

\bibitem{minguzzi} M. Amoruso, I. Meccoli, A. Minguzzi and M.P. Tosi, Eur. Phys. J. D {\bf 7}, 441 (1999).

\bibitem{griffin}  A. Griffin, Wen-Chin Wu and S. Stringari, Phys. Rev. Lett. {\bf 78}, 1838 (1997); Yu. Kagan, E.L. Surkov and G.V. Shlyapnikov, Phys. Rev. A {\bf 55}, R18 (1997).

\bibitem{kf} It is also  useful to introduce  the Fermi wave vector $k_F$ of the non interacting trapped Fermi gas, fixed by the relation $\hbar^2k_F^2/2M= (3N)^{1/3}\hbar\bar{\omega}$. In terms of $k_F$ one has  
 $ N^{1/6}a/a_{ho} = 0.59 k_Fa$.
 
\bibitem{vichi} L. Vichi and S. Stringari, Phys. Rev. {\bf 60}, 4734 (1999)

\end{thebibliography}
\end{document}